\documentclass[  aip, apl
 amsmath,amssymb,
 reprint,%
]{revtex4-1}

\usepackage{graphicx}
\usepackage{dcolumn}
\usepackage{bm}
\usepackage{xcolor}
\usepackage[utf8]{inputenc}
\usepackage[T1]{fontenc}
\usepackage{mathptmx}
\usepackage{etoolbox}
\makeatletter
\def\@email#1#2{%
 \endgroup
 \patchcmd{\titleblock@produce}
  {\frontmatter@RRAPformat}
  {\frontmatter@RRAPformat{\produce@RRAP{*#1\href{mailto:#2}{#2}}}\frontmatter@RRAPformat}
  {}{}
}%
\makeatother

\draft 

\begin{document}


\title{Electrically-induced resonance shifts of whispering gallery resonators made of barium magnesium fluoride} 



\author{Alexander Mrokon}
\affiliation{Laboratory for Optical Systems, Department of Microsystems Engineering - IMTEK, University of Freiburg, Georges-Köhler-Allee 102, 79110 Freiburg, Germany}
 \email{alexander.mrokon@imtek.uni-freiburg.de}
\author{Heike Kraft}
\affiliation{Laboratory for Optical Systems, Department of Microsystems Engineering - IMTEK, University of Freiburg, Georges-Köhler-Allee 102, 79110 Freiburg, Germany}
\author{Dongsung Shin}
\affiliation{Laboratory for Optical Systems, Department of Microsystems Engineering - IMTEK, University of Freiburg, Georges-Köhler-Allee 102, 79110 Freiburg, Germany}
\author{Hiroki Tanaka}
\affiliation{Leibniz-Institut für Kristallzüchtung (IKZ), Max-Born-Str. 2, 12489 Berlin, Germany}
\author{Simon J. Herr}
\affiliation{Fraunhofer Institute for Physical Measurement Techniques IPM, Georges-Köhler-Allee 301, 79110 Freiburg,Germany}
\author{Karsten Buse}
\affiliation{Laboratory for Optical Systems, Department of Microsystems Engineering - IMTEK, University of Freiburg, Georges-Köhler-Allee 102, 79110 Freiburg, Germany}
\affiliation{Fraunhofer Institute for Physical Measurement Techniques IPM, Georges-Köhler-Allee 301, 79110 Freiburg,Germany}
\author{Ingo Breunig}
\affiliation{Laboratory for Optical Systems, Department of Microsystems Engineering - IMTEK, University of Freiburg, Georges-Köhler-Allee 102, 79110 Freiburg, Germany}
\affiliation{Fraunhofer Institute for Physical Measurement Techniques IPM, Georges-Köhler-Allee 301, 79110 Freiburg,Germany}

\date{\today}

\begin{abstract}
Barium magnesium fluoride (BMF) is a ferroelectric crystal with a transparency range far beyond the one of other optical materials. In particular, its low loss in the deep ultraviolet makes this material an unique candidate for frequency conversion in this spectral range. Due to its relatively weak second-order nonlinearity, a resonant configuration such as an optical whispering gallery would be beneficial. We show that femtosecond-laser based material processing enables the reliable fabrication of BMF whispering gallery resonators with quality factors beyond $10^7$. Their resonance frequencies can be shifted linearly by applying electric fields between the $+c$ and $-c$ faces of the crystal. The slope of the shift is $-0.8$~MHz/(V/mm). It seems that the origin of this shift is piezoelectricity, while the electro-optic effect is negligible. Our results pave the way for millimeter-sized frequency converters in the deep ultraviolet. Furthermore, they indicate that a careful determination of fundamental material properties is still necessary.  
\end{abstract}


\maketitle 

\section{Introduction}
\label{sec:Introduction}
Barium magnesium fluoride (BaMgF$_4$, BMF) has been known for over 50 years \cite{Didomenico69} and has recently gained attention as a promising optical material due to its unique combination of properties. One of its most striking features is its exceptionally broad transparency range, extending from 130~nm wavelength in the deep ultraviolet to beyond 10~\textmu m wavelength in the mid-infrared.\cite{VilloraBirefringent-andQuasiPhase-Matching} This wide transparency, along with its non-centrosymmetric crystal structure (orthorhombic point group $mm2$), enables a variety of useful effects, including piezoelectricity,\cite{Recker74} the linear electro-optic effect, \cite{Recker74, Wang17} second- and third-order nonlinear optical responses, \cite{VilloraBirefringent-andQuasiPhase-Matching, ChenMeasurmentOfSecond-Order, ChenFemtosecondZ-Scan, Mateos14} and ferroelectricity, which allows for domain engineering.\cite{VilloraBirefringent-andQuasiPhase-Matching, Mateos14, Herr23} The combination of these properties makes BMF highly attractive for advanced optical applications.

However, the high transparency in the ultraviolet comes with a fundamental drawback: the second-order nonlinear response is relatively weak. The measured nonlinear coefficients range from 0.05~pm/V (Ref. \cite{Bergman75}) to 0.36~pm/V (Ref. \cite{ChenMeasurmentOfSecond-Order}) at a wavelength of 1064~nm, which is approximately two orders of magnitude smaller than those of lithium niobate. \cite{Seres01} This limitation can be mitigated by fabricating a monolithic, millimeter-sized whispering gallery resonator from the material, which enhances the power, confines the light and enables strong nonlinear interactions.\cite{Breunig16, Strekalov16} It has been demonstrated that such a resonator, even with a nonlinear coefficient as low as 0.075~pm/V, can achieve second-harmonic conversion efficiencies of a few percent using only milliwatt-level pump powers.\cite{Furst15} 

Despite its potential, manufacturing BMF-based whispering gallery resonators presents significant challenges. The material exhibits cleavage planes, which complicate precise shaping and polishing necessary for achieving high-quality optical modes.\cite{MinkovBMF} Additionally, controlling the eigenfrequencies of these resonators using an external electric field would be highly desirable for tunable photonic applications. A particularly notable characteristic of BMF is its strong piezoelectric response, which has been reported to be approximately four times as high as that of quartz.\cite{Recker74} However, its electro-optic response remains a topic of debate. While an early study suggest a weak response,\cite{Recker74} a recent one reports a value exceeding those of lithium niobate, one of the gold standards in electro-optic materials.\cite{Wang17} Due to this uncertainty, the feasibility of electrically-induced tuning remains unexplored.

In this contribution, we demonstrate a technique for fabricating high-quality BMF-based whispering gallery resonators, overcoming the cleavage-related challenges in material processing and achieving smooth, low-loss surfaces suitable for high-$Q$ resonances. Furthermore, we observe significant electrically induced resonance shifts indicating that the piezoelectric response is very large, while the electro-optic response plays only a minor role.

\section{Theoretical considerations}
\begin{figure*}[t]
\includegraphics{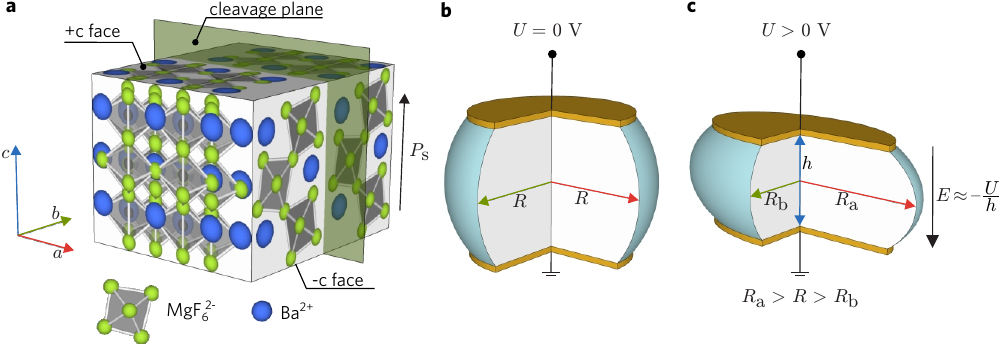}
\caption{\label{fig:Kristall} a) Atomic structure of barium magnesium fluoride (BaMgF$_4$), represented by projections in the $ab$, $ac$, and $bc$ planes. The spontaneous polarization $P_\mathrm{s}$ points from the $-c$ face to the $+c$ face of the structure. b) Rotationally symmetric whispering gallery resonator with zero voltage applied to the $+c$ face. The major radius in $a$ and $b$ directions is $R$. c) Whispering gallery resonator deformed by applying a positive voltage $U$ to the $+c$ face. The major radius in $a$ direction is increased to $R_\mathrm{a}$ while the major radius in $b$ direction reduces to $R_\mathrm{b}$. The electric field $E\approx -U/h$ points toward the $-c$ face, i.e., it is negative.}
\end{figure*}

We consider a BMF crystal oriented as sketched in Fig.~\ref{fig:Kristall}a. The $c$ axis is aligned such that its orientation is parallel to the spontaneous polarization $P_\mathrm{s}$ of the crystal. The $b$ axis is perpendicular to the cleavage plane, while the $a$ axis is perpendicular to $b$ and $c$. From this crystal, a whispering gallery resonator with the major radius $R$ is fabricated such that its rotation axis corresponds with the $c$ axis (see Fig. \ref{fig:Kristall}b). 

The aim of this work is to shift the resonance frequencies of a whispering gallery resonator by applying an electric field $E$ between its $+c$ and $-c$ faces. To first order, the resonance frequency $\nu$ is given by: \cite{Strekalov16}
\begin{equation}
    \nu=\frac{m c_0}{L n}\;,
\end{equation}
where $c_0$ is the speed of light in vacuum, $L$ is the geometric circumference of the resonator, $m$ denotes the number of oscillations in the equatorial plane, and $n$ is the refractive index of the bulk material. For millimeter-sized resonators, $m$ is on the order of 10,000, and the minor radius has only a negligible effect on the resonance frequency. \cite{Gorodetsky06}

A shift $\Delta\nu$ in the resonance frequency arises from changes in the resonator's circumference $\Delta L$ and refractive index $\Delta n$, leading to the expression:
\begin{equation}
    \frac{\Delta\nu}{\nu}=-\frac{\Delta L}{L}-\frac{\Delta n}{n}\,. \label{eq:shift}
\end{equation}

In the following, we analyze these two contributions separately, beginning with the effect of $\Delta L/L$. A change in the resonator’s circumference is induced by the converse piezoelectric effect. For BMF, the tensorial nature of this effect is given by
\begin{equation}
    \left(
\begin{array}{c}
\varepsilon_a\\
\varepsilon_b\\
\varepsilon_c\\
\varepsilon_{bc}\\
\varepsilon_{ac}\\
\varepsilon_{ab}
\end{array}
\right) = 
    \left(
\begin{array}{c c c}
0 & 0 & d_{31}\\
0 & 0 & d_{32}\\
0 & 0 & d_{33}\\
0 & d_{24} & 0\\
d_{15} & 0 & 0\\
0 & 0 & 0
\end{array}
\right)
    \left(
\begin{array}{c}
E_a\\
E_b\\
E_c\\
\end{array}
\right)\label{eq:piezotensor}
\end{equation}
Here, $\varepsilon_j$ represents the strain along the respective directions. The nonzero elements $d_{ij}$ have been determined as $d_{31}=-4.2$~pm/V, $d_{32}=2.5$~pm/V, $d_{33}=8.1$~pm/V, $d_{24}=-5.3$~pm/V, $d_{15}=-1.2$~pm/V (Ref. \cite{Recker74}). The electric field components are denoted by $E_i$. In our configuration, we assume a homogeneous electric field along the $-c$ direction, meaning $E_a=E_b=0$ and $E_c=-U/h$. Thus, we neglect relatively small changes in the electric field due to the curved rim of the resonator. \cite{Minet21} The relevant radii transform as $R_a=R(1-d_{31}U/h)$ and $R_b=R(1-d_{32}U/h)$. As discussed earlier, the deformation along the $c$-axis has only a minor impact on the resonance frequency. The resulting circumference $L$ is now elliptical, with semiaxes $R_a$ and $R_b$ where $R_a>R>R_b$ (see Fig.~\ref{fig:Kristall}c). Since the eccentricity remains small, we approximate $L=2\pi R(1-(d_{31}+d_{32})U/(2h))$ (Ref. \cite{bronstein1985taschenbuch})and consequently
\begin{equation}
    \frac{\Delta L}{L}=-\frac{d_{31}+d_{32}}{2}\frac{U}{h}
\end{equation}
with the effective coefficient $(d_{31}-d_{32})/2=-0.85$~pm/V. For $U=100$~V and $h=1$~mm, we estimate a relative circumference change of $8.5\times10^{-8}$ which corresponds to a resonance-frequency shift of $\Delta\nu=-25$~MHz for light at 1~\textmu m wavelength. Importantly, this shift is independent of the light polarization.

Next, we examine the contribution of $\Delta n/n$ in Eq. (\ref{eq:shift}), with the piezoelectric coefficients $d_{ij}$ replaced by the electro-optic coefficients $r_{ij}$ in Eq. (\ref{eq:piezotensor}) and strain components $\varepsilon_j$ replaced by changes in impermeability $\Delta\eta_j$, from which the refractive-index change is determined as $\Delta n_j = -n_j^3\Delta\eta_j/2$. The coefficients $r_{ij}$ remain largely unknown for BMF. While Recker et al. reported only a weak electro-optic response \cite{Recker74} compared with the piezoelectric one without specification, Wang et al. have measured $r_{31}=-36.2$~pm/V (Ref. \cite{Wang17})


For light polarized along the $c$ axis, the relative refractive-index shift is $\Delta n_c=n_c^3r_{33}U/(2h)$. For light polarized perpendicular to the $c$ axis, the refractive indices $n_a$ and $n_b$ change as $\Delta n_a = n_a^3r_{13}U/(2h)$ and $\Delta n_b = n_b^3r_{23}U/(2h)$. Since the refractive index varies along the roundtrip path, the resonance frequency is determined by the average refractive index $\bar{n}=(n_a+n_b)/2$ (Ref. \cite{Furst16}) leading to
\begin{equation}
   \frac{\Delta n}{n} =
\left\{
\begin{array}{ll}
\frac{1}{2}n_c^2r_{33}\frac{U}{h}, &\quad \text{Polarization $\parallel$ c} \\
 & \\
\frac{1}{2}\bar{n}^2\frac{r_{31}+r_{32}}{2}\frac{U}{h}, &\quad \text{Polarization $\perp$ c}
\end{array}
\right.
\end{equation}
\begin{figure*}[t]
\includegraphics[width=\textwidth]{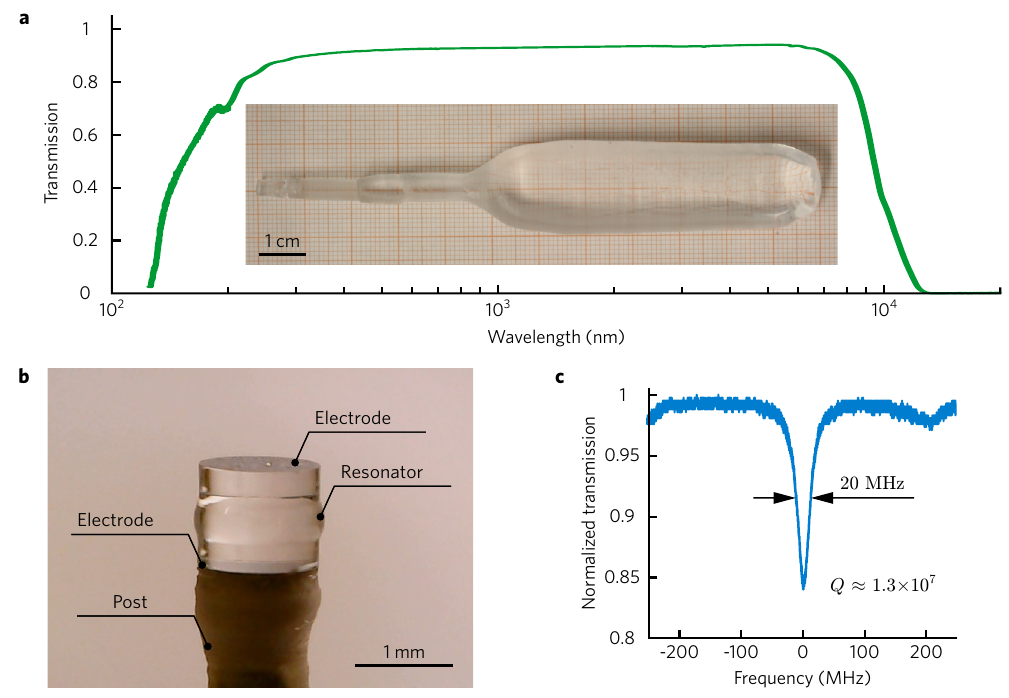}
\caption{\label{fig:Quality} a) Photograph of a barium magnesium fluoride (BMF) boule and transmission spectrum of a 1.8~mm thick sample. b) Photograph of a whispering gallery resonator (WGR) made of BMF. c) Transmission spectrum of across a whispering gallery resonance with the corresponding quality factor $Q$ at 1064~nm wavelength.}
\end{figure*}
Since only the value for $r_{31}$ was determined experimentally, a quantitative estimation of the electro-optic response is not possible. However, if it is significant compared with the piezoelectric one, the frequency shift will strongly depend on light's polarization. This dependence would only vanish in the unlikely scenario where $r_{33}\approx(r_{13}+r_{23})/2$.

\section{Resonator fabrication}
\label{sec:ResFab}

The resonators used in this study were fabricated from a high-purity boule grown at the Leibniz Institute for Crystal Growth (IKZ).\cite{Herr23} 
The BMF crystal was grown along the crystal's c-axis by the Czochralski technique from a stoichiometric melt. At a  growth speed of 0.8 mm/h, a crack-free boule was obtained.
Figure~\ref{fig:Quality}a presents a photograph and a transmission spectrum of a 1.8-mm-thick crystal cut from this boule. As expected, the crystal is transparent from the deep ultraviolet to the mid-infrared. The boule is sliced into chips with the dimensions $(10\times20\times1)~\text{mm}^3$ in $a$, $b$ and $c$ directions, respectively. The $+c$ and $-c$ faces of the chips are coated with a 150-nm-thick chromium layer deposited via sputtering.

Traditionally, whispering gallery resonators are fabricated as follows \cite{GrudininUltrahighOpticalQ}: A cylindrical preform is extracted from the chip using a hollow drill. This preform is then shaped with a computer-controlled lathe equipped with a diamond knife. Finally, the surface is polished using progressively finer abrasives. This procedure is well established in many laboratories and has been successfully applied to a wide range of materials. However, it is not suitable for reliably fabricating whispering gallery resonators from BMF. In particular, the use of a hollow drill and a diamond knife causes significant cracks along the cleavage plane perpendicular to the crystal's $b$-axis ((010) plane). Recently, it was demonstrated that a fabrication process exceeding 100 hours, involving various abrasives followed by final chemo-mechanical polishing, can produce optical whispering gallery resonators from BMF with a quality factor of approximately $10^7$ at 1550~nm wavelength.\cite{MinkovBMF}

\begin{figure*}[t]
\includegraphics{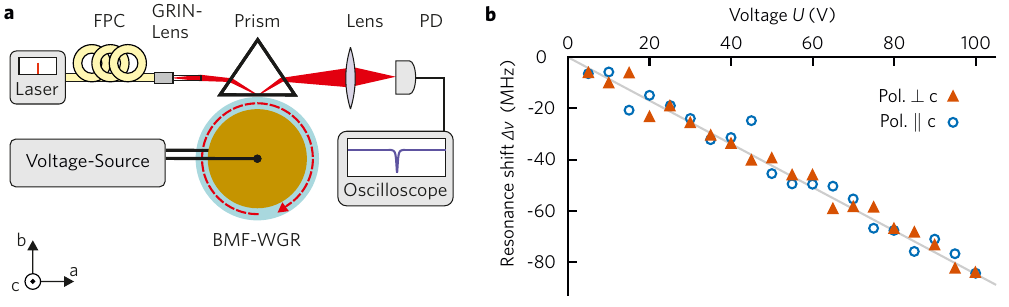}
\caption{\label{fig:Setup} a) Experimental setup for measuring the WGR resonance shift under an applied DC voltage, including a fiber polarization controller (FPC) and photodiode (PD). b) Measured resonance shift as a function of applied voltage. A maximum frequency shift of 80~MHz is observed at 100~V. The tuning rate is determined to be 0.8~MHz/(V/mm) for both polarization states. A grey line is included as a visual guide.
}
\end{figure*}

We introduced an alternative approach for fabricating whispering gallery resonators. \cite{WernerGeometricTuning} In this method, the critical steps — cutting out the preform and shaping the resonator rim — are performed using 200-fs laser pulses at a wavelength of 388 nm. This process significantly reduces the risk of damage associated with the material’s cleavage planes. Subsequent polishing is carried out with a diamond slurry containing 1-µm particles to remove surface contaminants from the shaping process, followed by a final polishing step using 50-nm particles. The entire polishing procedure takes well under an hour. Using this procedure, millimeter-sized whispering gallery resonators can be reliably fabricated. The resonator shown in Fig.~\ref{fig:Quality}b has a major radius of 0.7~mm and a minor radius of 0.3~mm, while the distance between the electrodes is $h=1$~mm. 

To determine the quality factor, we measure the transmission spectrum across a whispering gallery resonance at 1064~nm wavelength (see Fig.~\ref{fig:Quality}c). In the undercoupled regime, the linewidth $\delta\nu$ is approximately 20~MHz, resulting in a quality factor of around $1.3\times10^7$. This value is consistent with that observed using the 100-hour polishing procedure. From the linewidth, we estimate the propagation loss as $\alpha=2\pi n\delta\nu/c0~\approx6\times10^{-3}$/cm (Ref. \cite{LeidingerComparativeStudy}) Assuming that surface scattering plays only a minor role, this value primarily reflects the extinction from the bulk material, indicating that the bulk material is of high quality.

\section{Voltage-Controlled Resonance Shift}
\label{sec:ElectricalTuning}

To determine the electrically-induced shifts of the resonance frequencies, the whispering gallery resonator is integrated into the experimental setup shown in Fig.\ref{fig:Setup}a. Laser light at 1064~nm wavelength, with a linewidth of 10 kHz and a 1~mW power, is coupled into the resonator via frustrated total internal reflection using a prism made of SF11 glass. To enable the application of electric fields, a DC voltage source is connected to electrodes positioned on the $+c$ and $-c$ faces of the resonator. The light exiting the prism is focused onto a photodetector, which is linked to an oscilloscope for monitoring the transmission spectrum of the resonator, as shown in Fig~\ref{fig:Quality}c. Transmission spectra are recorded for voltages ranging from 0 to 100~V and for light polarization both parallel and perpendicular to the $c$ axis.

Fig.~\ref{fig:Setup}b shows the experimental results. By applying a positive voltage to the $+c$ face of the resonator, the resonance frequencies linearly shift to smaller values with the slope $\Delta\nu/U = -0.8$~MHz/V. Furthermore, this shift does not depend on the polarization of light. 

Now, we compare these observations with the theoretical predictions above. We recall that resonance shifts due to the converse piezoelectric effect are predicted with $-0.25$~MHz/V slope independent of light's polarization. The magnitude of electro-optically induced resonance shifts is unclear. However, a significant polarization dependence can be expected. Our experimental observations align well with the theoretical prediction of resonance shifts by the converse piezoelectric effect in terms of sign, order of magnitude, and polarization characteristics. However, the absolute value of the measured slope is approximately three times larger than expected. The effective coefficient $(d_{31}-d_{32})/2$ would be $-2.7$~pm/V rather than $-0.85$~pm/V as expected, i.e. it would correspond approximately to $d_{11}$ of quartz. The missing polarization dependence indicates that the electro-optic response plays only a minor role. This general behavior is consistent with results of earlier experiments.\cite{Recker74}


\begin{figure*}[t]
\includegraphics{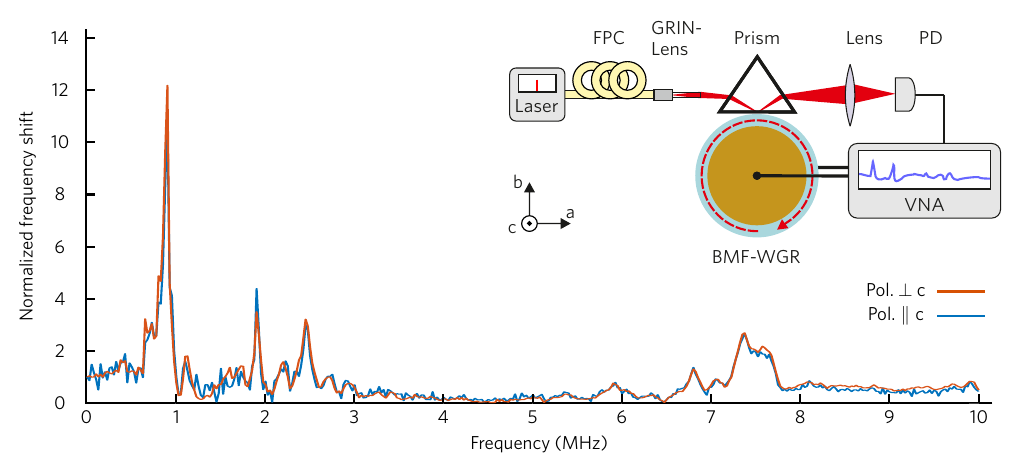}
\caption{\label{fig:EOS21} Frequency dependence of the resonance shift. Increased resonance shifts are found at 0.91, 1.93, 2.46, and 7.35~MHz. Inset: Schematic of the measurement setup, including a fiber polarization controller (FPC), photodiode (PD), and vector network analyzer (VNA).
}
\end{figure*}

We further support this by determining the frequency dependence of the resonance shifts using the setup displayed in Fig.~\ref{fig:EOS21}. The laser frequency is red-detuned from the resonance frequency to the value that gives the highest slope of the transmission signal. In Fig.~\ref{fig:Quality}c, it would be $+10$~MHz detuning from the resonance. Port 1 of a vector network analyzer (VNA) is used to apply sinusoidal voltage signals from 9~kHz to 10~MHz. The temporally varying voltage translates to a temporally varying intensity measured by the photodiode. It's output is fed into the network analyzer's port 2. Using this scheme, the normalized frequency shift as a function of excitation frequency is determnined. \cite{HongTangHighFrequency} This is done for light polarization both parallel and perpendicular to the $c$ axis.

Figure~\ref{fig:EOS21} shows the experimental results. It is obvious that the AC response of the resonance shift is independent of light's polarization. At 908~kHz excitation frequency, the shift of the resonance frequency exceeds the DC value by a factor of twelve. Also at 1.932, 2.456, and 7.352~MHz, we observe a significant increase in the electrically induced eigenfrequency shift, presumably due to excitation of the resonator's mechanical resonances.

The missing polarization dependence in the whole range between 9~kHz and 10~MHz indicates that the shift of the eigenfrequencies in our experiment is just due to the change of the geometrical circumference. That means the converse piezoelectric effect is by far the dominant contribution.
However, its experimentally determined magnitude is more than three times larger than predicted. This indicates that the piezoelectric coefficients of the crystals used in our experiment do not match the ones determined in the early study.\cite{Recker74} Thus, it seems that even basic physical properties of BMF still have to be determined carefully with recently grown high-quality crystals.

\section{Conclusion}
\label{sec:Conclusion}

Barium magnesium fluoride is so-far unmatched among non-centrosymmetric crystals regarding its transparency range, particularly in the deep ultraviolet. Due to its relatively small second-order nonlinearity, resonant configurations are essential to achieve highly efficient frequency conversion under continuous-wave excitation. We have demonstrated that femtosecond-laser-assisted manufacturing enables the reliable fabrication of millimeter-sized whispering gallery resonators with a quality factor exceeding $10^7$, despite the raw material's propensity to cleave perpendicular to the crystallographic $b$ axis. Furthermore, the eigenfrequencies can be controlled electrically.

We consider this an important step towards realization of miniaturized and efficient frequency converters in the deep ultraviolet, based on both second-order and third-order nonlinearity. In particular, the combination of monolithic resonators with domain engineering \cite{VilloraBirefringent-andQuasiPhase-Matching, Mateos14, Herr23} is expected to significantly advance this development.

Our experiments indicate that the electrically-induced shift of the eigenfrequencies is predominantly driven by the converse piezoelectric effect. However, the measured magnitude is three times larger than predicted. This discrepancy suggests that the previously determined coefficients of the piezoelectric tensor\cite{Recker74} may not accurately reflect those of the recently grown high-quality material. Additionally, there remains uncertainty regarding the electro-optic response. While our experiments, as well as earlier work,\cite{Recker74} indicate a minor effect compared to the piezoelectric response, another study reports a significantly stronger electro-optic effect.\cite{Wang17} Consequently, a comprehensive characterization of the raw material's fundamental properties is still needed.

The pronounced frequency dependence of the electrically-induced shift of the eigenfrequencies is a particularly intriguing feature. At specific excitation frequencies, this shift is enhanced by more than an order of magnitude, which could might be beneficial for highly efficient optomechanical modulation in integrated photonic circuits.\cite{HongTangSub-Terahertz}

We believe that our work is a first step for integrating barium magnesium fluoride into photonic circuits, offering new opportunities for optomechanical and nonlinear-optical applications.


\begin{acknowledgments}
We acknowledge support by the Open Access Publication Fund of the University of Freiburg.
This work was financially supported by the German Research Foundation, DFG (BR 4194/12-1).

\end{acknowledgments}

\section*{Data Availability Statement}

The data that support the findings of this study are available from the corresponding author upon reasonable request.



%
%

%


\bibliography{BMF}

\end{document}